\documentclass[aps,reprint,groupedaddress,superscriptaddress]{revtex4-1}
\usepackage{amsmath}
\usepackage{amssymb}
\usepackage{amsfonts}
\usepackage[pdftex]{graphicx}
\usepackage{textgreek}
\usepackage{makecell,tabularx}
\setcellgapes{3pt}

\begin{document}

\title{Deformation of a rotated granular pile governed by body-force-dependent friction}

\author{Terunori Irie}
\affiliation{Department of Earth and Environmental Sciences, Nagoya University, Furocho, Chikusa, Nagoya 464-8601, Japan}

\author{Ryusei Yamaguchi}
\affiliation{Technical Center, Nagoya University, Furocho, Chikusa, Nagoya 464-8601, Japan}

\author{Sei-ichiro Watanabe}
\affiliation{Department of Earth and Environmental Sciences, Nagoya University, Furocho, Chikusa, Nagoya 464-8601, Japan}

\author{Hiroaki Katsuragi}
\affiliation{Department of Earth and Space Science, Osaka University, 1-1 Machikaneyama, Toyonaka 560-0043, Japan}

\date{\today}

\begin{abstract}
  Although the gravity dependence of granular friction is crucial to understand various natural phenomena, its precise characterization is difficult. We propose a method to characterize granular friction under various gravity (body force) conditions controlled by centrifugal force; specifically, the deformation of a rotated granular pile was measured. To understand the mechanics governing the observed nontrivial deformation of this pile, we introduced an analytic model considering local force balance. The excellent agreement between the experimental data and theoretical model suggests that the deformation is simply governed by the net body force (sum of gravity and centrifugal force) and friction angle. The body-force dependence of granular friction was precisely measured from the experimental results. The results reveal that the grain shape affects the degree of body-force dependence of the granular friction.  
\end{abstract}

\maketitle

\section{Introduction}

Friction is one of the simplest but most complex features that results in various nonlinear behaviors of granular matter~\cite{Duran:2000,Midi:2001,Andreotti:2013}. Particularly, when considering astronomical situations such as the terrain dynamics of terrestrial bodies covered with regolith, the gravity dependence of granular friction is the most crucial factor. However, the accurate measurement of the gravity dependence of granular friction is difficult. 
Many studies have attempted to reveal the gravity dependence of the granular angle of repose, which is determined by friction, by using rotating drums~\cite{Klein:1990,Arndt:2006,Cosby:2009,Kleinhans:2011} or other methods~\cite{Hofmeister:2009,Blum:2010,Marshall:2018,Chen:2019}; however, they obtained rather contradictory results. Some studies reported a positive correlation between the angle of repose and gravitational acceleration, while others obtained the opposite tendency or almost constant behaviors~\cite{ Klein:1990, Cosby:2009,Kleinhans:2011, Hofmeister:2009,Marshall:2018,Chen:2019}. Because the gravity dependence of the granular angle of repose is weak, very careful measurements are necessary for accurately characterizing it. 

The localization of flow (also known as shear banding) in granular matter ~\cite{Lemieux:2000,Komatsu:2001,duPont:2005,Katsuragi:2010} is a universal characteristic that could also depend on gravity. The thickness of the localized flow depends on the experimental setup and conditions~\cite{Jop:2005,Tsuji:2019,Arndt:2006}. The accurate measurement of the gravity-dependent flow thickness has also been difficult. These granular behaviors, friction and flow localization, are crucial for analyzing the surface terrains of astronomical bodies covered with regolith/boulders~\cite{Melosh:2011,Katsuragi:2016,Walsh:2018}. In particular, the gravity dependence of these quantities in a quasi-static regime is a key issue for discussing diverse planetary phenomena that gradually proceed over a long period. Therefore, we focus on the friction and local flow thickness of granular matter in the quasi-static regime under various body-force (gravity) conditions. 

However, the characterization of static granular friction is more difficult than that of dynamic granular friction. In a dynamically flowing regime, granular friction can be characterized by an empirical law called $\mu-I$ rheology~\cite{daCruz:2005,Jop:2006,Pouliquen:2006}. However, granular static states exhibit strong protocol dependence~\cite{Vanel:1999,Bertho:2003}. Even the static granular friction angle shows complexity~\cite{Nagel:1992}. To overcome the difficulties in the static characterization of granular matter, we develop an experimental apparatus that enables the accurate measurement of the quasi-static deformation of a granular pile~\cite{Irie:2021}. Using the developed apparatus, we perform a set of experiments to measure the net gravity (body force) dependence of quasi-static granular friction. The surface flow thickness on the deforming granular pile is also measured. To explain the gravity-dependent granular behavior, a simple force-balance model is introduced. Using the analyzed results, we discuss the relation between the granular friction, net body force due to gravity and centrifugal force, and surface flow thickness. We also discuss a possible implication for asteroidal shape development.

%%%%%%%%%%%%%%%%%%%%%%%%%%%%%%%%%%%%%%%%%%%%%%%%%%%%%%%%%%%%%%%%%%%%%%%%%%%%%%%
\begin{figure*}[hbt]
\begin{center}
\includegraphics[width=18cm]{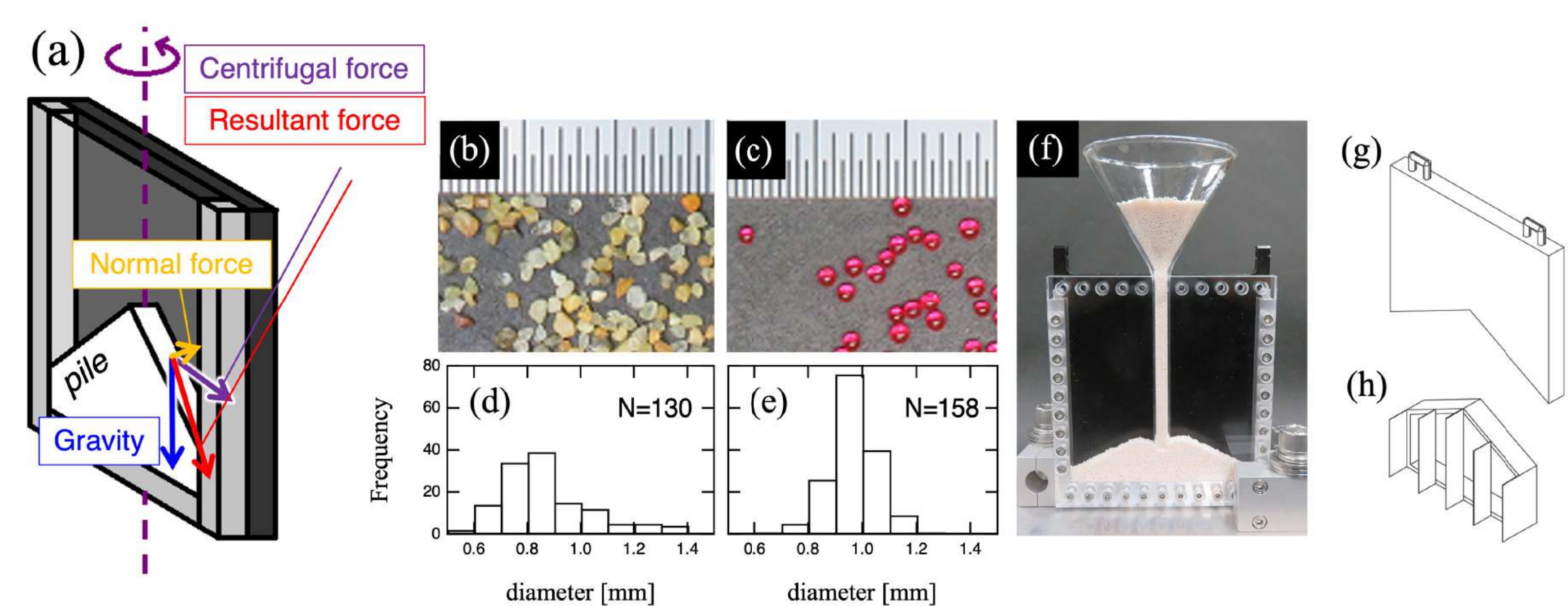}
    \caption{Experimental setup, used grains, and preparation methods. (a)~Schematic of the setup. A granular pile is rotated around the vertical axis. Shape of (b)~sand grains and (c)~glass beads, and size distribution of (d)~sand and (e)~glass beads are presented. The grain size was defined by the simple mean of the lengths of the major and minor axes of the approximated ellipsoid shape of each grain. (f)~The sand pile was prepared using a funnel. However, a stripe-patterned glass-bead pile cannot be formed by the funnel method. Instead, we used two molds: (g)~spacer mold and (h)~partitioner mold. By placing these molds on a horizontally laid cell, a stripe-patterned glass-bead pile was prepared. Then, the molds were removed. 
}
\label{fig:apparatus}
\end{center}
\end{figure*}
%%%%%%%%%%%%%%%%%%%%%%%%%%%%%%%%%%%%%%%%%%%%%%%%%%%%%%%%%%%%%%%%%%%%%%%%%%%%%%%%

\section{experiment}

In the experiment, a quasi-two-dimensional granular pile consisting of sand or glass beads was rotated around the vertical axis as shown in Fig.~\ref{fig:apparatus}(a). To observe the deformation of the rotated granular pile, we developed an experimental apparatus that simultaneously rotates both the granular pile and a camera unit (Raspberry Pi 3 with ArducamSKU:B0032)~\cite{Irie:2021}. We used sand (paddy sand, AF Japan) and colored glass beads (Ballotini) to form the granular piles. The average grain sizes $d$ of sand and glass beads were 0.9~mm and 1.0~mm, respectively. Both grains had an identical true density $\rho_t=2.5 \times 10^3$~kg~m$^{-3}$ and initial bulk density (true density times packing fraction) $\rho_b=1.4 \times 10^3$~kg~m$^{-3}$. Actual photos of the grains and corresponding size distributions are shown in Fig.~\ref{fig:apparatus}(b-e). 

The granular piles were formed in a quasi-two-dimensional cell with the inner dimensions of 100~mm $\times$ 100~mm $\times$ 10~mm. The initial sand pile with an angle of repose $\theta_r$ was formed by a funnel (Fig.~\ref{fig:apparatus}(f)). However, to form the striped glass-bead pile, we used molds made of acrylic resin that partitioned the pile into six columns. The initial glass-bead pile was prepared as follows. (i) The molds were set on a horizontally laid cell. (ii) The areas were separated by a partitioner mold (Fig.~\ref{fig:apparatus}(h)) were separately filled with colored glass beads. (iii) The partitioner mold was pulled out, and the sample cell was closed. Then, (iv) the sample cell was raised vertically, and (v) the spacer mold that regulated the surface shape (Fig.~\ref{fig:apparatus}(g)) was pulled out. The slope of the spacer mold was fixed to the angle of repose of the glass beads. We employed a granular pile with the angle of repose as an initial condition so that the quasi-static deformation of the pile could be observed from the beginning of deformation when the centrifugal force was gradually increased from zero. 

After constructing a granular pile in the cell, the cell was mounted at the center of the rotating unit. Then, the rotation rate was gradually increased so that the dimensionless centrifugal effect $\Gamma=r_0 \omega^2/g$ varied from $0.02$ to $21.5$, where $r_0=50$~mm, $\omega$, and $g=9.8$~m~s$^{-2}$ are the radius (half width) of the cell, rotational angular speed, and gravitational acceleration, respectively. $\Gamma$ was increased in a stepwise manner. The maximum rotation rate was $620$~rpm. We confirmed that the equilibrium shape was achieved for each rotation rate. Therefore, the deformation proceeded in a quasi-static manner when the increase in the rotation rate was less than or equal to the current condition, $1$~rpm~s$^{-1}$. The equilibrium shape was captured by the camera unit. The acquired images were analyzed (binarized) to identify the surface profiles (see Appendix~A for the image analysis method). Three experimental runs were conducted under identical conditions to verify the reproducibility. Details of the experimental apparatus and procedures are provided in Ref.~\cite{Irie:2021}. 

\section{Results and Analyses}

%%%%%%%%%%%%%%%%%%%%%%%%%%%%%%%%%%%%%%%%%%%%%%%%%%%%%%%%%%%%%%%%%%%%%%%%%%%%%%%
\begin{figure*}[hbt]
\begin{center}
\includegraphics[width=18cm]{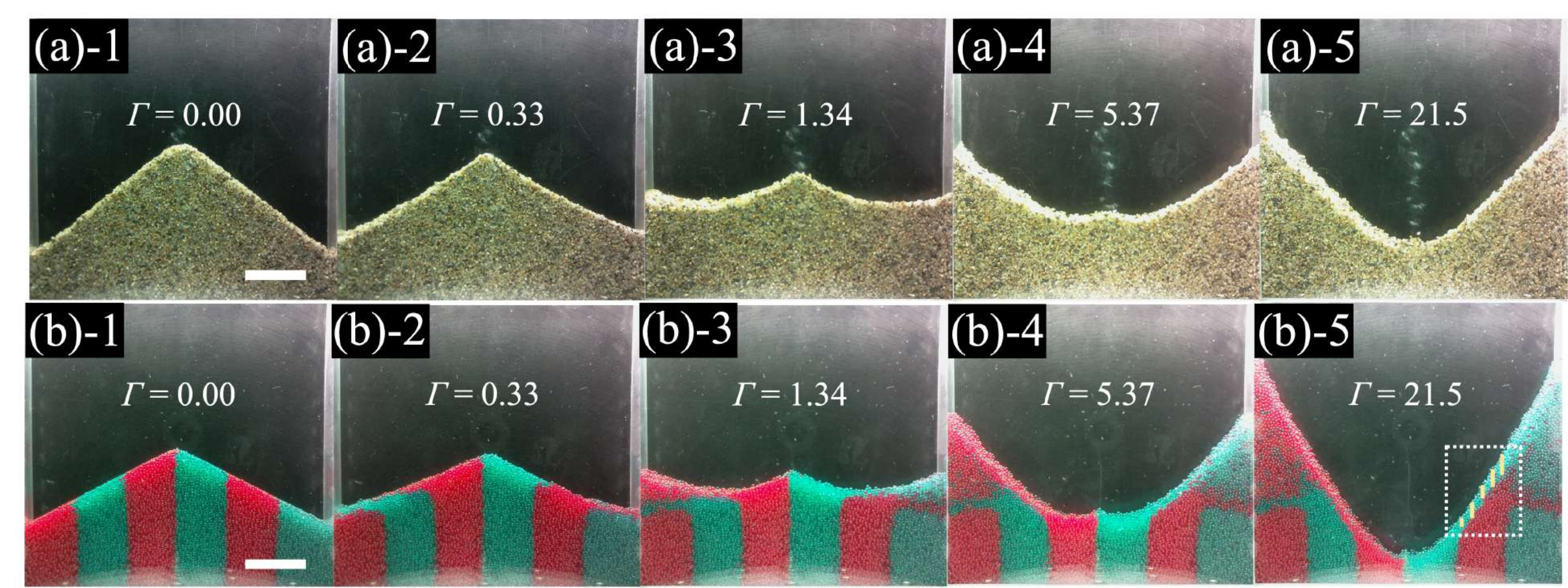}
    \caption{Experimental raw data. (a) Sequential deformation of a sand pile. Corresponding $\Gamma$ values are presented in each panel. The gradual relaxation and steep-angle development on the side walls are observed as $\Gamma$ increases. (b) Sequential deformation of a colored (stripe-patterned) glass-bead pile. Qualitative behavior of the glass-bead pile is similar to that of the sand pile shown in~(a). Localized surface flow is confirmed in (b). The surface flow thickness $\delta$ was measured at five positions in the middle portion of the cell~(white box in (b)-5). Scale bars indicate 20~mm. %Videos of granular pile deformation are provided in Supplemental Material. 
}
\label{fig:raw_data}
\end{center}
\end{figure*}
%%%%%%%%%%%%%%%%%%%%%%%%%%%%%%%%%%%%%%%%%%%%%%%%%%%%%%%%%%%%%%%%%%%%%%%%%%%%%%%%

\subsection{Surface profile}
The measured examples of the deforming granular piles consisting of sand or colored glass beads are shown in Fig.~\ref{fig:raw_data}(a) and (b), respectively. The initial shape of the granular pile shows a constant slope with an angle of repose $\theta_r$. As $\Gamma$ increases, the granular pile gradually relaxed, and the grains were transported from the central to the outward region. As shown in Fig.~\ref{fig:raw_data}(b), the stripe pattern of the colored glass beads was used to visualize the flowing state in the deforming granular pile. The green and red parts consist of glass beads that are identical except for the color. As shown in Fig.~\ref{fig:raw_data}(b), the flowing region is localized in the thin surface layer, implying that a typical shear-banding structure can be observed in a rotated granular pile as well.  

\begin{figure*}[hbt]
\begin{center}
\includegraphics[width=18cm]{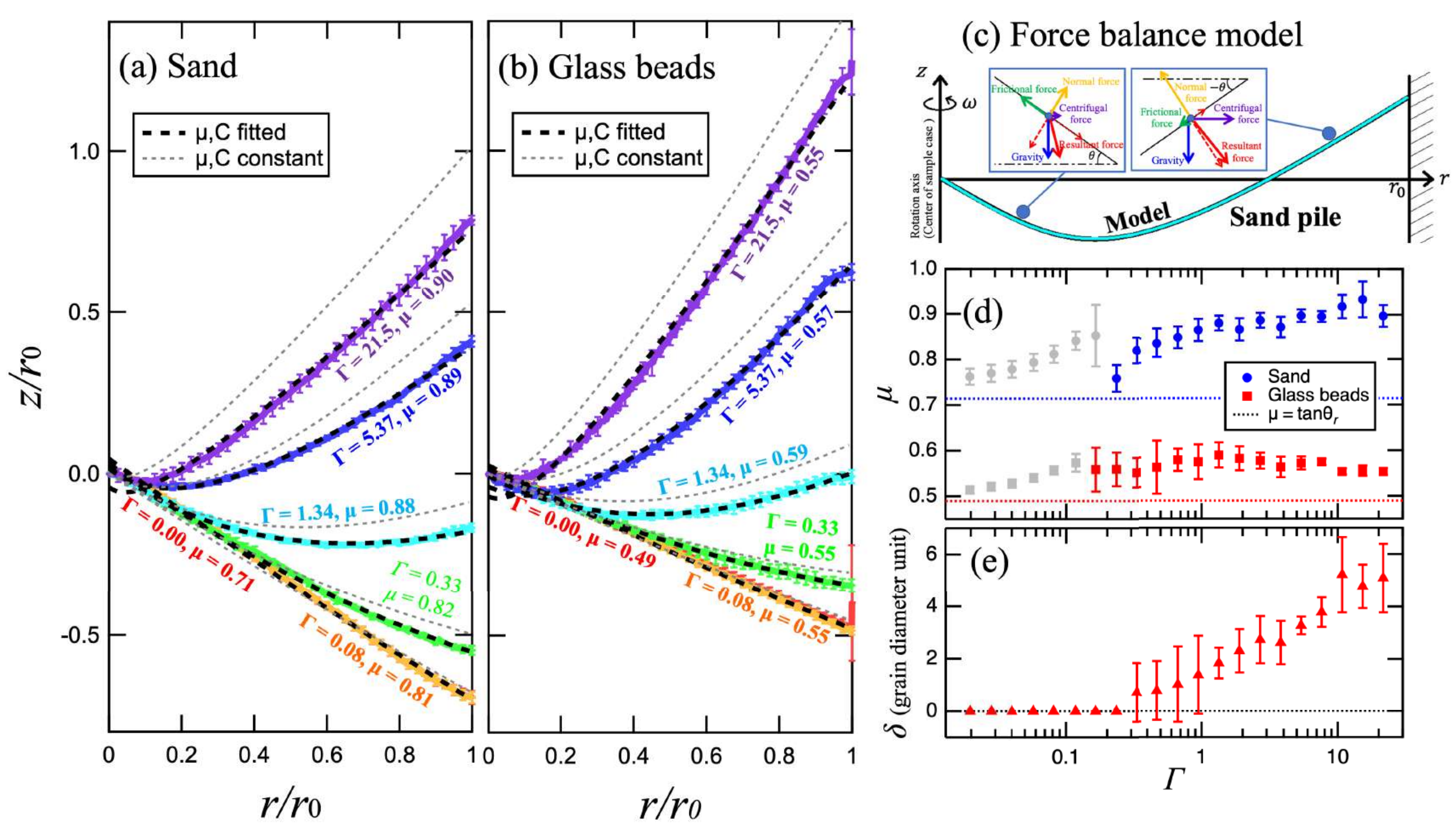}
    \caption{Surface profiles of the deforming granular pile and the analyzed results. In panels (a) and (b), colored curves with error bars show the experimental results. Profiles of the (a)~sand pile and (b)~colored-glass-bead pile are presented. The thick dashed curves in black are the fitting by Eq.~\eqref{eq:model_sol} with fitted $\mu$ and $C$ values. Corresponding $\Gamma$ and fitted $\mu$ values are provided beside the data. All the experimental results can be well fitted by the model curves. The thin dotted curves indicate the constant $\mu$ and $C$ curves ($\mu=0.71$ for sand, $\mu=0.49$ for glass beads, and $C=0$ for both grains), where $\mu$ is fixed by the angle of repose $\mu=\tan \theta_r$. The constant $\mu$ and $C$ curves deviate from the data particularly in a large-$\Gamma$ regime. Error bars indicate the standard deviation of six profiles (left and right of three experimental runs). (c) Schematic of the force-balance model of Eq.~\eqref{eq:model_dif_eq}. Balance among three forces (gravity, centrifugal force, and friction) is simply considered through the resultant and normal forces. This balance should be satisfied everywhere on the surface profiles. (d)~$\mu(\Gamma)$ obtained by the fitting. The increasing trend of $\mu(\Gamma)$ can be confirmed particularly in sand pile deformation. Gray symbols indicate that the granular pile was not deformed. Horizontal dotted lines in (d) indicate the friction coefficient corresponding to the angle of repose, $\mu=\tan\theta_r$. (e)~$\delta(\Gamma)$ measured by the deformation of colored (stripe) glass-bead pile. The $\delta(\Gamma)$ function exhibits an increasing trend. 
}
\label{fig:profiles}
\end{center}
\end{figure*}

First, we analyze the surface profiles governed by granular friction. 
The surface profiles identified via image analysis are shown in Fig.~\ref{fig:profiles}(a) and (b). Owing to the axisymmetric rotation, the height of profile $z$ should be a function of the horizontal distance from the rotation axis $r$. Thus, Fig.~\ref{fig:profiles}(a) and (b) show the average $z/r_0$ profiles as functions of $r/r_0$. The height $z$ and distance $r$ are normalized to the cell width $r_0$, and the height at $r=0$ is fixed as $z(0)=0$ to clearly separate the profiles. While both piles show qualitatively similar deformation, the observed shapes represent nontrivial complex curves.

%%%%%%%%%%%%%%%%%%%%%%%%%%%%%%%%%%%%%%%%%%%%%%%%%%%%%%%%%%%%%%%%%%%%%%%%%%%%%%%
%%%%%%%%%%%%%%%%%%%%%%%%%%%%%%%%%%%%%%%%%%%%%%%%%%%%%%%%%%%%%%%%%%%%%%%%%%%%%%%%

To understand the deformation mechanics of the rotated granular pile, we propose a simple model. As shown in Fig.~\ref{fig:profiles}(c), the force balance among gravity, friction, and centrifugal force is considered along the direction of the slope in the rotating frame of reference. Although the former two forces can be considered constant, the centrifugal force depends on $r$. The local angle $\theta$ (clockwise direction is defined as positive, as shown in Fig.~\ref{fig:profiles}(c)) should be determined to balance the three forces. The force balance along the slope direction (per unit mass) is written as
\begin{equation}
g \sin\theta + r\omega^2 \cos\theta  = \mu \left[ g \cos\theta - r\omega^2 \sin\theta \right],
\label{eq:force_balance}
\end{equation} 
where, $\mu$ is the friction coefficient. By using the geometric condition $\tan\theta=-dz/dr$, Eq.~\eqref{eq:force_balance} is rewritten as
\begin{equation}
\frac{dz}{dr} = \frac{\left(r\omega^2/g\right) -\mu}{1+\mu\left(r\omega^2/g\right)}.
\label{eq:raw_dif_eq}
\end{equation}

If the system is frictionless ($\mu=0$), Eq.~\eqref{eq:raw_dif_eq} becomes $dz/dr= r\omega^2/g$. Then, a parabolic shape is obtained with $z(r)= (\omega^2/2g)r^2$. This parabolic solution well reproduces the rotated water surface deformation ~\cite{Irie:2021}. The current model (Eq.~\eqref{eq:raw_dif_eq}) additionally considers the finite $\mu$ case. The friction effect tends to stabilize the surface profile. 
From the relation $\Gamma=r_0 \omega^2/g$, the local slope $dz/dr$ should obey the following equation:
\begin{equation}
\frac{dz}{dr}= \frac{\Gamma r^* -\mu}{1+\mu \Gamma r^*},
\label{eq:model_dif_eq}
\end{equation}
where $r^*=r/r_0$ is the normalized distance. At the initial state ($\Gamma=0$), the equation becomes $\tan\theta=dz/dr=-\mu$ where the negative sign originates from the geometrical definition. At the opposite limit ($\Gamma=\infty$), this equation can be simplified as $dz/dr=1/\mu$. These limiting cases correspond to the angle of repose relative to the bottom and vertical sidewall, respectively. Therefore, the granular heap with an angle of repose is appropriate as an initial configuration to confirm the validity of the model. In other words, this initial condition naturally satisfies Eq.~\eqref{eq:model_dif_eq} at all places in the granular pile. As $\Gamma$ increases from $\Gamma=0$, the local slope varies depending on $\Gamma$, $r$, and $\mu$. That is, this model corresponds to a simple extension of the definition of the friction angle. 

Equation~\eqref{eq:model_dif_eq} can easily be integrated as 
\begin{equation}
  \frac{z}{r_0} =\frac{1}{\mu}\left[ r^* - \left( \frac{1}{\mu\Gamma} + \frac{\mu}{\Gamma} \right) \ln \left(\mu\Gamma r^* + 1 \right) \right]+C.  
\label{eq:model_sol}
\end{equation}
Here, $C$ is an integration constant. 

In this model, we assume a continuum-like behavior of the granular surface by considering the surface force balance. We do not consider the microscopic contacts among the grains and the resultant complex internal stress distribution in the granular pile. The $\mu$ measured in this study represents the macroscopic friction coefficient governing the free-surface deformation. For the static case ($\Gamma=0$), a similar idea has been used to characterize granular friction~\cite{Nedderman:1992,Fayed:1997}. It should be noted that this friction coefficient is not necessarily identical to the friction among grains. 

To obtain this model, $\Gamma$ and $r_0$ are determined by the experimental conditions, and $\mu$ is a material parameter. Therefore, the experimental results should ideally agree with the model curves, without any free fitting parameter. To verify this, the curves with constant $\mu(=\tan\theta_r)$ and $C=0$ are shown as thin dotted curves in Fig.~\ref{fig:profiles}(a) and (b). Although the curves capture the qualitative trend of the surface profiles, the large $\Gamma$ profiles cannot be reproduced by these curves. This deviation indicates the weak but finite gravity dependence of granular friction. In addition, while we fix $C=z(0)=0$ to separately show all the experimental profiles, the central part ($r\simeq 0$) is not significantly affected by the centrifugal force. We considered the uncertainty of $C=z(0)$ in the analysis. Therefore, we treated $\mu$ and $C$ as fitting parameters. The resulting fitting curves (thick dashed curves) are in excellent agreement with the experimental results (Fig.~\ref{fig:profiles}(a) and (b)). Owing to the accurate measurement of the profiles and the excellent fitting results, we confirm that a weak $\Gamma$ dependence of $\mu$ has been detected. The value of $C$ only affects the vertical level of the curves, and the fitted $C$ values are on the order of $10^{-3}$--$10^{-2}$; the effect of $C$ is minor. Thus, we focus on the variation of $\mu$ in this study.

\subsection{Body-force-dependent granular friction}
The obtained relation between $\mu$ and $\Gamma$ is shown in Fig.~\ref{fig:profiles}(d). As shown, $\mu$ is an increasing function of $\Gamma$ for sand pile deformation. When $\Gamma$ is less than $0.15$ (grey symbols in Fig.~\ref{fig:profiles}(d)), the granular pile was not deformed owing to the effective cohesion strength~\cite{Irie:2021}. The model of Eq.~\eqref{eq:model_sol} should not be applied in this regime, because it does not include the cohesion effect. The possible sources of the cohesion are van der Waals forces, capillary effects, and electrostatic effects~\cite{Nagaashi:2021,Kimura:2015,Blanc:2011,Bocquet:1998}. The cohesion effect may be related to the difference between the angle of repose and starting angle of a granular slope. Refer to \cite{Irie:2021} for the detailed analysis and discussion of the cohesion strength based on the deformation of the rotated granular pile. Moreover, the horizontally rotated granular cylinder was also used for the characterization of the cohesion strength~\cite{Herminghaus:2013}. However, in this study, we focus on the continuous deformation governed by the body force and friction, rather than the cohesive characteristic. Once the deformation is triggered for large values of $\Gamma$($>0.2$), significant deformation of granular piles is observed in every step. In this regime, the effect of cohesion is negligible as the centrifugal force overwhelms the cohesion force. Namely, the angle of repose was measured and analyzed in this study. The excellent agreement between the experimental results and model curves with fitted $\mu$ values indicates that the local slope is determined by the local resultant body-force direction (sum of gravity and centrifugal force) and gravity-dependent friction angle. In other words, the free-surface deformation is induced by the surface force balance rather than the internal stress distribution. 
The dotted lines shown in Fig.~\ref{fig:profiles}(d) indicate the level of initial angle of repose, $\mu=\tan\theta_r$. All the measured $\mu$ values are greater than the initial values. However, the increasing trend of $\mu(\Gamma)$ is more evident for sand piles than that for glass-bead piles. This result suggests that the degree of gravity (body-force) dependence of granular friction depends on the grain shape. 
Because the deformation is induced by the surface flow (Fig.~\ref{fig:profiles}(c)), we next focus on the surface flow characterization. 
 
\subsection{Surface flow thickness}
As shown in Fig.~\ref{fig:raw_data}(b), relatively thin surface flow dominates the deformation. To further characterize the surface flow, we analyzed the development of the cumulative flow thickness at the middle part of the granular pile. 
By using the vertically striped glass-bead pile (Fig.~\ref{fig:raw_data}(b)), the thickness of the surface flow $\delta$ that is gradually developed during the deformation can be measured. The flow thickness to transport grains from the central region to the outside region $\delta$ was measured in the middle part of the right half of the striped glass-bead pile (white box in Fig.~\ref{fig:raw_data}(b-5)). The thickness of green beads on the red column was measured at five positions, as shown in Fig.~\ref{fig:raw_data}(b-5). To evaluate the loading-history dependence of the surface flow thickness, $\delta$ was measured in all $\Gamma$ states. In Fig.~\ref{fig:profiles}(e), the average of $\delta$ measured at five positions is displayed in the unit of grain diameter $d$. A clear increasing trend of $\delta(\Gamma)$ can be confirmed because the flow thickness is accumulated during the deformation. This behavior is different from the usual shear banding observed in avalanching~\cite{Lemieux:2000,Komatsu:2001} or vibro-fluidization~\cite{Tsuji:2019} wherein the flow thickness is constant or proportional to the height. According to Fig.~\ref{fig:profiles}(e), the thickness of $\delta$ is at most $5d$--$6d$ at $\Gamma=21.5$. In the large $\Gamma$ regime, the flow thickness $\delta$ was large because a substantial number of grains were transported from the central region toward the outside region. This cumulative deformation under the large body-force condition effectively strengthens the granular friction. In other words, the grain network connectivity is strengthened owing to the network rearrangement induced by the large inclined (shearing) body force or the increase in the body force. When $\Gamma$ was small, $\delta$ was almost zero due to the small but finite cohesive effect~\cite{Irie:2021}. 

\section{Discussion}
Friction is typically characterized by the ratio of shear and normal forces. In this study, we applied this conventional definition of $\mu$ to the additional body force (centrifugal force) environment. Geometrically, the angle between the local slope and local horizon (perpendicular to the resultant body-force direction) represents the friction angle, $\arctan(\mu)$. Furthermore, $\Gamma$ indicates the degree of body-force loading. Thus, the geometrical analysis of the profile under various $\Gamma$ states serves as an appropriate approach to reveal the sole body-force effect in granular static friction through a simple extension of the conventional definition of $\mu$. The body-force loading and surface-force loading do not necessarily result in an identical outcome. Indeed, the current experimental result indicates a nontrivial $\mu(\Gamma)$ relation. However, in this study, the body-force dependence of friction is demonstrated for the static granular deformation alone. Further studies on dynamic granular deformation (or flow) and frictional behaviors of bulk materials under the effect of centrifugal force are necessary. Furthermore, $\mu$ is only a phenomenological parameter that characterizes granular friction. The $\mu$ value must depend on the microscopic features of the granular force distribution etc. In this study, however, such details cannot be considered. The detailed study of microscopic features governing $\mu$ variation is an important problem that will be considered in future work.  

A possible idea explaining the increasing $\mu(\Gamma)$ is the increase in packing fraction $\phi$. In a sheared granular layer, the internal friction coefficient (as a ratio of shear and normal stresses) and packing fraction (or equivalently void ratio) can be utilized to evaluate the failure condition~\cite{Roscoe:1958,Chen:1998}. In addition, an empirical relation between friction coefficient and void ratio can be obtained by a shear test~\cite{Fayed:1997}. Namely, $\phi$ could be a crucial parameter to analyze $\mu$ behavior. If this is the case for the current situation, the increase in packing fraction $\phi$ induced by the additional body force (centrifugal force) is the main reason for the increase in $\mu(\Gamma)$. However, it is difficult to confirm the systematic relation between $\mu$ and $\phi$. Specifically, the sand pile shows a positive relation between $\mu$ and $\phi$, whereas the glass-bead pile shows a weak negative correlation (see Appendix~B). 
Thus, it is difficult to conclude that $\mu(\phi)$ relation is the main reason for the increase in $\mu(\Gamma)$. In this study, the value of $\phi$ was measured for the entire granular heap (Appendix~B). 
However, the actual deformation is driven by the local surface flow. Thus, the local packing fraction should be analyzed to precisely discuss the effect of $\phi$ on $\mu$.

On a steady rapid granular heap flow confined in a quasi-two-dimensional cell, the effective slope could be greater than that of the static state due to the wall friction effect~\cite{Taberlet:2003}. The value of $\mu$ (slope) measured in any quasi-two-dimensional experiment (including this study) certainly includes the effect of wall friction. However, the centrifugal force in this experiment does not affect the normal loading to the walls. In addition, the deformation of the granular pile proceeded quasi-statically, and the flow rate was sufficiently small to neglect the inertial effect of the flowing grains. Nevertheless, we observed the increasing trend of $\mu$ by increasing $\Gamma$. Although the specific value of $\mu$ depends on the boundary conditions such as the gap width, the increasing behavior of $\mu$ should be qualitatively independent of the boundary conditions. To quantify the wall effect, experiments with various thickness cells and/or three-dimensional cells should be performed. This problem can be addressed in future work.

If the rotation rate is decreased after its increase, the initial shape of the granular pile cannot be recovered. This implies that the deformation is irreversible. The origin of this irreversibility is the effect of friction. Due to the nonlinear stability caused by friction, the deformation induced by the increase in the rotation rate can be maintained when the rotation rate is slightly reduced. Thus, the observed history dependence can be well understood by considering the friction effect. In this study, we focused on the increasing $\Gamma$ regime for simplicity. The history-dependent deformation can be analyzed by slightly expanding the force-balance model. The quantitative investigation of the history-dependent behavior is an important problem that could be discussed in a future study.

A similar experimental study was reported recently (after the submission of this paper)~\cite{Huang:2021}. The study also analyzed the quasi-static deformation of the rotated granular pile to discuss the asteroidal surface process. The local-slope distribution and its variation by the centrifugal force were carefully analyzed. Moreover, an analytic surface-profile model was derived by assuming the constant angle of repose relative to the horizon defined by the local body force direction. This idea is effectively identical to our force balance model. Indeed, Eq.~(2) in \cite{Huang:2021} and Eq.~(\ref{eq:model_sol}) in this paper are mathematically equivalent. The study reported that the local angle ranges from the angle of repose to the starting angle (dynamic and static angles). However, in this study, we successfully fit the entire profile by the theoretical model. The principal advantages of our study, compared to \cite{Huang:2021}, are as follows. (i)~The initial shape (angle of repose) is a critically stable state. (Huang et al. employed the horizontal initial condition~\cite{Huang:2021}.) Thus, we can observe the quasi-static deformation of the entire granular pile. (ii)~The maximum rotation rate in this study (620~rpm) is faster than that in \cite{Huang:2021} (300~rpm). (iii)~Surface flow was visualized by rotating a stripe-patterned pile. (iv)~Furthermore, the physical explanation of the theoretical model is more detailed in this study. Based on these advantages, we revealed the body-force-dependent granular friction in this study. We successfully reproduced all profiles using the theoretical model with a single $\mu$ value. This is in contrast to the study reported by Huang et al., wherein the inner and outer profiles can be separately fitted by different curves~\cite{Huang:2021}. Moreover, we found that the body-force dependence of $\mu$ could be affected by the grain shape. Further investigation of the granular deformation under the effect of centrifugal force is necessary to fully reveal its physical nature. 

Finally, we briefly discuss the planetological implication of our experimental results. The top shape of the small asteroid Ryugu can be formed by the effect of the rotational centrifugal force and self-gravity. The experimental results indicate that the local shape of the granular surface is simply determined by the local direction of the net gravity (sum of centrifugal force and gravity) and friction. This is consistent with the observation of Ryugu. Watanabe et al. reported that Ryugu's top shape can be explained by a rapid rotation period of 3.5~h (current rotation period is 7.6~h) and a roughly constant angle of repose of 31$^{\circ}$~\cite{Watanabe:2019}. That is, its top shape could have been formed by ancient rapid rotation, and the shape has been preserved by the hysteresis effect of the granular slope. This situation is similar to the early stage of the current experiment. Our experimental results provide supportive evidence for this. They show that a shallow surface flow driven by centrifugal force deforms into the asteroidal shape. While the variation of $\mu(\Gamma)$ is not very significant, its quantitative effect on the asteroidal-shape development is unclear. The hysteresis effect of $\mu$ is also a crucial factor for various applications. Detailed characterization of the $\mu(\Gamma)$ variation and hysteresis effect is open to future. Furthermore, the mixing of the spectrally different surface granular materials may be induced by the surface flows on the asteroid Ryugu~\cite{Morota:2020}. The experimental setup shown in Fig.~\ref{fig:raw_data}(b) may be beneficial in quantitatively analyzing the mixing process in such granular surface flows. 

While the direction of gravitation is almost constant in the laboratory setup, it is location-dependent in the asteroidal environment. In this study, we successfully constructed a model explaining the granular deformation under the constant gravity condition. To consider the actual asteroidal situation, the current model must be extended to the case of arbitrary gravity direction in three-dimensional space. Such a generalization of the model is an interesting open problem. The current experimental result and model provide a fundamental basis for further investigations on asteroidal global shape determination and its surface processes.

\section{Conclusion}
We performed systematic experiments and analyses of granular-pile deformation using the centrifugal effect. The deformed surface profile of the granular piles can be well explained by the local force balance model. After data fitting, the increase in the friction coefficient $\mu$ was confirmed by increasing the centrifugal force, particularly for the sand-pile deformation. Cumulative surface flow thickness $\delta$ was also measured, and its positive correlation to the centrifugal effect $\Gamma$ was confirmed. A thin but growing surface flow might affect the frictional state of the centrifuged granular pile. Although the gravity (body force) dependence of granular friction has been long debated, our experimental results and theoretical model clearly reveal this dependence. However, the increasing tendency of $\mu(\Gamma)$ could depend on the grain shape. Further investigation is necessary to fully understand the effect of gravity on various types of granular behaviors.

\begin{acknowledgments}
SW and HK thank JSPS KAKENHI for financial support under Grant No.~19H01951, 18H03679, and 17H06459.
\end{acknowledgments}
  
\appendix
\section{Three-dimensional effect of the surface profile (depth inclination)}
\label{sec:sup_depth_incli}

\begin{figure}
\begin{center}
\includegraphics[width=8cm]{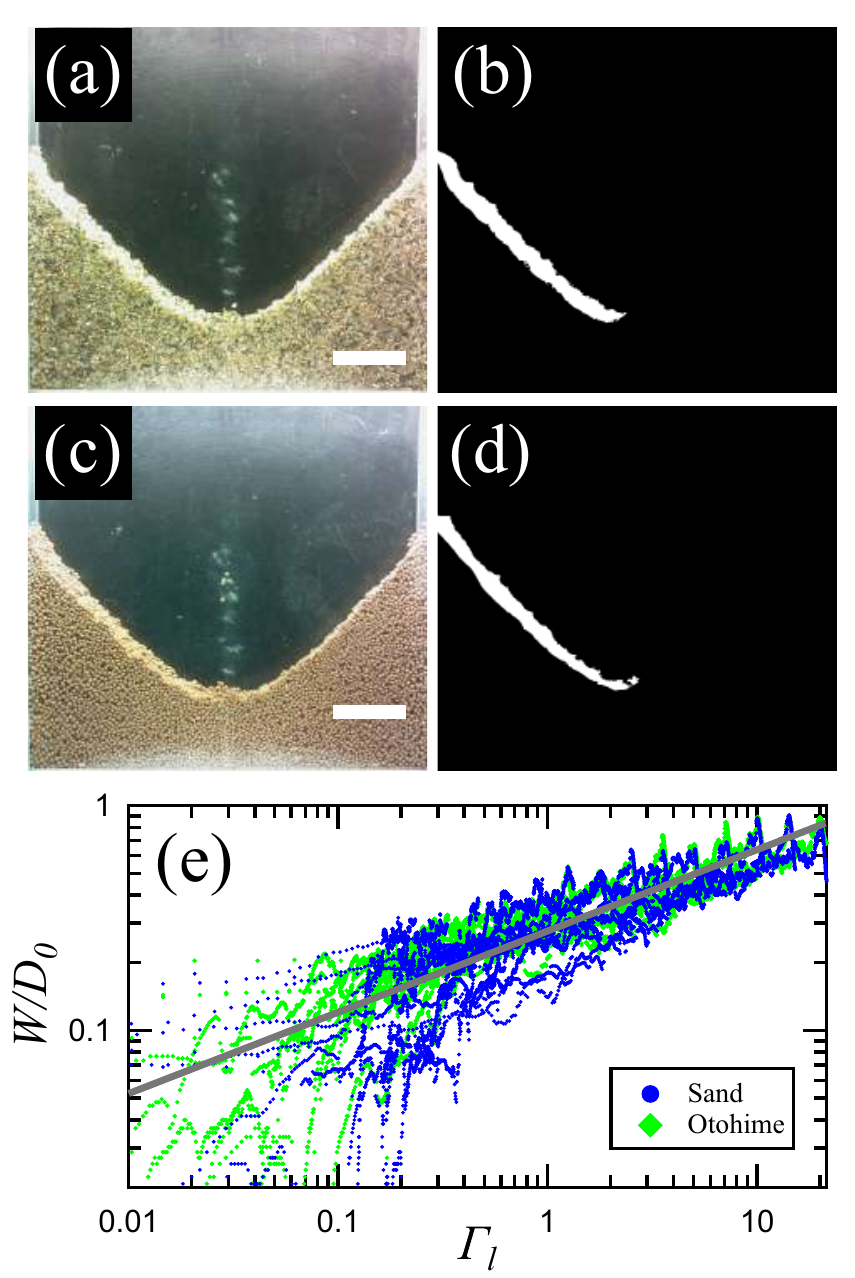}
    \caption{Depth-inclination effect. As shown in panels (a) and (c), grains on the surface of the left side develop an inclination to the cell's depth direction. To obtain a reasonable mean height, we extract the depth-inclination band through image analysis, as shown in panels (b) and (d). Then, the center of the depth-inclination band is traced as a mean surface height. This effect is clearly observed in the sand pile~(a). However, it cannot be identified in the glass-bead pile. Therefore, depth inclination produced on the surface of a spherical-grain pile was measured using opaque and dark-color grains (fish feed called Otohime), as exemplified in (c) and (d). Scale bars indicate 20~mm. (e)~Measured depth-inclination width versus local centrifuge factor $\Gamma_l$. Because both data follow an identical trend, we use a fitting function (Eq.~\eqref{eq:depth_inclination}) to correct for the glass-bead pile.
}
\label{fig:depth}
\end{center}
\end{figure}

The three-dimensional effect of the finite depth of the cell ($D_0=10$~mm) should be considered to properly estimate the packing fraction $\phi$. Because we used grains of $d\simeq 1$~mm, the thickness in the depth direction is approximately $10d$. By accelerating the rotation, the grains move outwards owing to the centrifugal force. Simultaneously, the grains move towards the depth direction owing to the inertial effect. This effect could incline the surface profile along the depth direction. Indeed, this effect is visible in some actual data of the sand pile, as shown in Fig.~\ref{fig:depth}(a). Note that the granular pile was illuminated from above. 
Hereinafter, we call this the three-dimensional inclination the ``depth inclination.'' Although this effect does not significantly affect the shape of profiles, it is crucial for appropriately measuring the volume of the granular pile. In this analysis, we assumed linearity in the depth inclination. Then, the profile at the center of the depth-inclination band should correspond to the mean height. To specifically compute the representative (mean) height, we extracted the depth-inclination band (bright part) from the raw data, as shown in Fig.~\ref{fig:depth}(b). This inclination is visible only in the left half because the inertial effect exhibits the opposite tendency in the right half. From the depth-inclination band data shown in Fig.~\ref{fig:depth}(b), we can compute the mean height profiles both in the left and right halves by assuming that the system is axisymmetric. However, the depth inclination can be clearly observed only for sand. It is difficult to identify the depth inclination for the spherical glass beads directly from the raw data owing to certain technical reasons ~(e.g., the brightness is too high and/or the grains are transparent). To obtain the general law for the depth-inclination effect, we plot the normalized thickness of depth inclination $W/D_0$ as a function of the local $\Gamma$ in Fig.~\ref{fig:depth}(e). Here, $W$ is the measured thickness of the depth-inclination band, $D_0=10$~mm is depth of the cell, and the local $\Gamma$ is defined as $\Gamma_l=(r\omega^2)/g$ (local distance $r$ is used instead of $r_0$). As shown in Fig.~\ref{fig:depth}(e), the thickness of the depth-inclination band shows a universal dependence on $\Gamma_l$. To verify the universality of this relation, we also used fish-feed grains called ``Otohime,'' with the properties of $d=1.1$~mm, $\rho_t=1.1 \times 10^{3}$~kg~m$^{-3}$, and $\rho_b=0.61 \times 10^3$~kg~m$^{-3}$. Because the grains are opaque, the depth inclination can clearly be observed in the Otohime piles (Fig.~\ref{fig:depth}(c) and (d)). The Otohime grains are porous and roughly spherical, while the sand grains possess a rough shape and dense structure. Nevertheless, they both obey the same universal relation, as shown in Fig.~\ref{fig:depth}(e). Therefore, we assume this relation is applicable to glass beads as well. Specifically, we empirically assume the relation 
\begin{equation}
\frac{W}{D_0} =A \Gamma_l^{\alpha},
\label{eq:depth_inclination}
\end{equation}
where $A = 2.8\times 10^{-2}$ and $\alpha=0.36$ are fitting parameters. By using Eq.~\eqref{eq:depth_inclination}, the profiles of the colored glass beads can also be corrected. 

%%%%%%%%%%%%%%%%%%%%%%%%%%%%%%%%%%%%%%%%%%%%%%%%%%%%%%%%%%%%%%%%%%%%%%%%%%%%%%%
%%%%%%%%%%%%%%%%%%%%%%%%%%%%%%%%%%%%%%%%%%%%%%%%%%%%%%%%%%%%%%%%%%%%%%%%%%%%%%%%

\section{Correction of $\mu$ and $\phi$}
\label{sec:sup_mu_phi}

\begin{figure}
\begin{center}
\includegraphics[width=8cm]{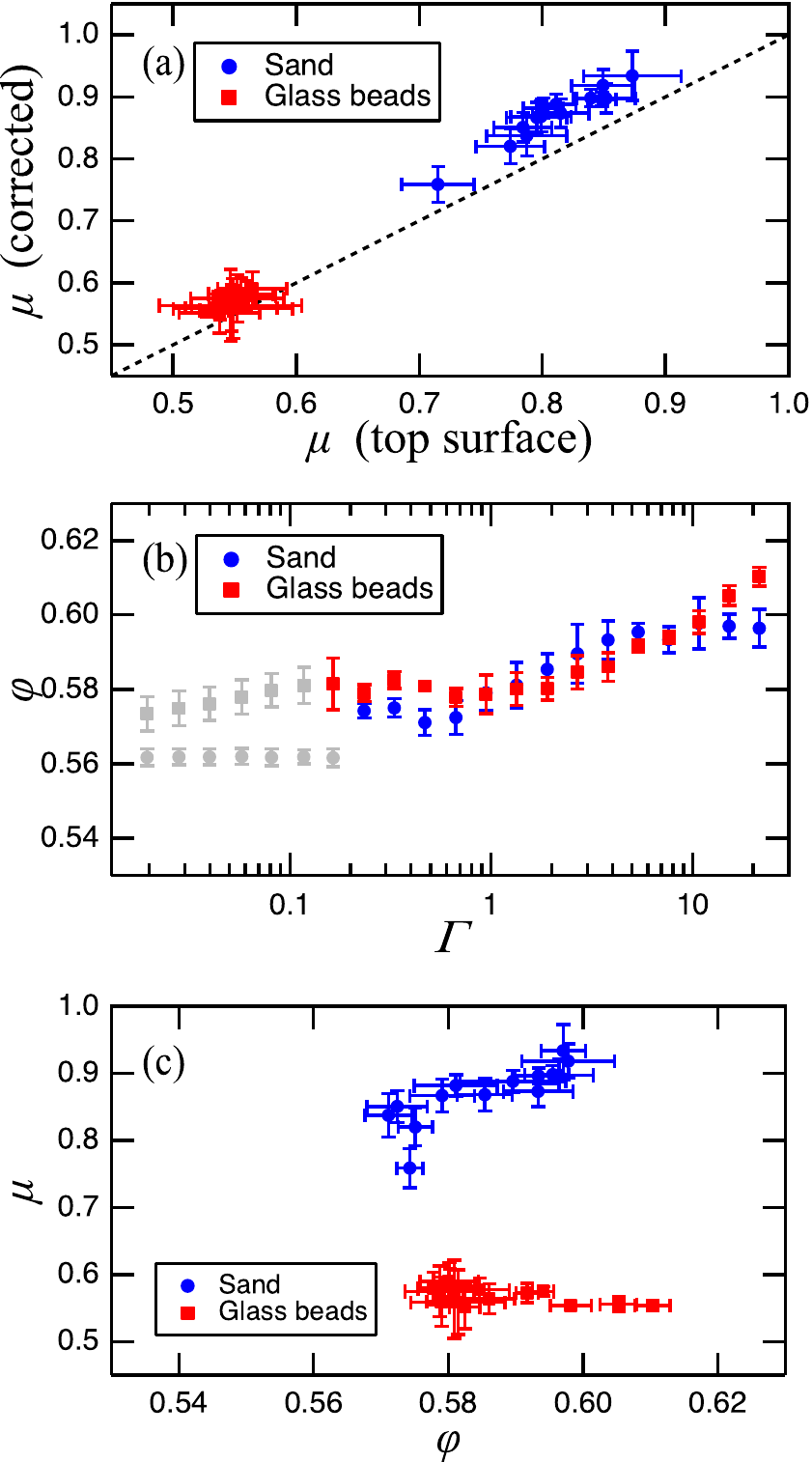}
    \caption{Values of $\mu$ and $\phi$ computed from the mean height. (a) Comparison of $\mu$ obtained by fitting to the corrected mean height and $\mu$ obtained by fitting to the top surface of the profile. The dashed line corresponds to the relation (corrected $\mu$) = (top surface $\mu$). Although the corrected $\mu$ shows slightly larger values, the difference is not significant. (b) Packing fraction $\phi$ as a function of $\Gamma$. To obtain a reasonable $\phi$, the correction of the depth inclination is necessary. An increasing $\phi(\Gamma)$ trend can be observed. Grey symbols indicate that the granular pile was not deformed. (c) Correlation between $\mu$ and $\phi$. The correlation manner depends on the grain shape.
    }
\label{fig:corrected}
\end{center}
\end{figure}

The value of $\mu$ is not very sensitive to the absolute height of the surface profile. Basically, the local slope is essential. When we employed the mean corrected profiles as actual surfaces, the value of $\mu$ determined by fitting will be slightly affected. A comparison of $\mu$ obtained by the corrected mean profile (vertical axis) and that obtained using the top surfaces (horizontal axis) is shown in Fig.~\ref{fig:corrected}(a). If the data lie on the dotted line, both $\mu$ values are the same. As shown in Fig.~\ref{fig:corrected}(a), a systematic difference can be observed. However, good linearity implies that the qualitative behaviors are not affected by the correction. Therefore, we used the mean corrected profiles for the analysis. 

The finite-size effect of the cell including the sidewall effect could affect the measured $\mu$ value. Because we fixed the cell size, it is difficult to assess the size effect. The experimental result with water could be explained by using a similar theoretical model without any fitting parameter~[Irie et al., submitted (2021)]. This result suggests that the effects of the system size and wall friction are almost negligible. To reveal the effect of system size, we should rotate a larger cell; this is not very easy to perform now owing to technical limitations. The improvement of the experimental apparatus is a future problem.

%%%%%%%%%%%%%%%%%%%%%%%%%%%%%%%%%%%%%%%%%%%%%%%%%%%%%%%%%%%%%%%%%%%%%%%%%%%%%%%
%%%%%%%%%%%%%%%%%%%%%%%%%%%%%%%%%%%%%%%%%%%%%%%%%%%%%%%%%%%%%%%%%%%%%%%%%%%%%%%

The correction is considerably more important for the computation of the packing fraction $\phi$ because the packing fraction is inversely proportional to the volume of the sand pile. In the absence of this correction, the volume was overestimated. Consequently, an unrealistic dilation of the granular pile (over 10\%) was estimated. By correcting the surface height, we obtained a reasonable $\phi(\Gamma)$ relation, as shown in Fig.~\ref{fig:corrected}(b). Both the sand and glass beads showed an increasing trend. This means that the compaction could be induced by applying the centrifugal force. Because the resultant body force due to the sum of the centrifugal force and gravity is greater than gravity alone, the compaction of the granular pile can be induced. The increasing $\mu(\Gamma)$ could relate to the increase in $\phi(\Gamma)$. Indeed, a certain correlation between $\mu$ and $\phi$ was found in a sheared granular matter~(Chapter 4 in \cite{Fayed:1997}).

The relation between $\mu$ and $\phi$ is shown in Fig.~\ref{fig:corrected}(c). For sand, we can confirm the positive correlation between $\mu$ and $\phi$. However, the negative correlation can be observed for glass beads. Thus, it is difficult to confirm a clear relationship between $\mu$ and $\phi$. Moreover, note that the $\phi$ shown in Fig.~\ref{fig:corrected}(b) is the \textit{bulk} packing fraction. The compaction could be induced mainly on the surface region as shown in Fig.~\ref{fig:raw_data}(b). The accurate \textit{local} measurement of $\phi$ is necessary for discussing the relation between the compaction and friction in detail. Such a focused measurement is a future problem.

\bibliography{rot_sandpile}

\end{document}